\documentclass[reprint,showpacs,floatfix,amsmath,amssymb,prb]{revtex4-2}
\usepackage{graphicx,tikz,latexsym,hyperref,colortbl}
\usepackage{multirow} \usepackage{mathrsfs}
\usepgflibrary{shapes.arrows}
\usepgfmodule{decorations}
\usepackage{fancyhdr}
\usepackage{ragged2e}
\usepackage{textcomp}
\newcommand{\fref}[1]{Figure~\ref{fig:#1}}
\newcommand{\tref}[1]{Table~\ref{tab:#1}}

\newcommand{\ket}[1]{\big|#1\big\rangle}             
\hypersetup{colorlinks=false, citecolor=purple, linkcolor=red}
\begin{document}
\title{  The  Jahn-Teller active  fluoroperovskites  $A\mathrm{CrF_3}$
  $A=\mathrm{Na^+},\mathrm{K^+}$:   thermo-    and   magneto   optical
  correlations as function of the $A$-site}

\author{F.~L.~M. Bernal}
\author{F.~Lundvall}
\author{S.~Kumar}
\author{P-A.~Hansen}
\author{D.~S. Wragg}
\author{H.~Fjellv\aa g}
\affiliation{Centre for Material Science and Nanotechnology, University of Oslo, NO-0315, Norway}
\author{O.~M. L\o vvik}
\affiliation{Department of Physics, University of Oslo, NO-0315, Norway}
\affiliation{Sintef Industry, 0314 Oslo, Norway}

\begin{abstract}
  Chromium                    (II)                   fluoroperovskites
  $A\mathrm{CrF_3}(A\mathrm{=Na^+,K^+})$   are   strongly   correlated
  Jahn-Teller active materials at low temperatures.  
  In this  paper, we examine the  role that the $A$-site  ion plays in
  this family of fluoroperovskites  using both
  experimental  methods  (XRD,  optical  absorption  spectroscopy  and
  magnetic fields) and DFT simulations.
  Temperature-dependent optical  absorption experiments show  that the
  spin-allowed transitions  $E_2$ and  $E_3$ only merge  completely 
   for $A$= Na at  2 K.
   Field-dependent optical  absorption measurements  at 2 K  show that
   the  oscillating  strength  of   the  spin-allowed  transitions  in
   $\mathrm{NaCrF_3}$ increases with increasing applied field.
   Direct magneto-structural correlations which suppress the spin-flip
   transitions  are  observed  for  ${\rm KCrF_3}$  below  its  Ne\'el
   temperature.  In  ${\rm NaCrF_3}$ the spin-flip  transitions vanish
   abruptly  below  9  K revealing  magneto-optical  correlations  not
   linked to crystal structure changes. This suggests that as the long
   range  ordering  is reduced  local  JT  effects in  the  individual
   ${\rm CrF_6^{4-}}$ octahedra take control of the observed behavior.
  Our  results show  clear deviation  from the  pattern found  for the
  isoelectronic $A_x{\rm MnF}_{3+x}$ system.  The size of the $A$-site
  cation is shown  to be central in dictating  the physical properties
  and phase transitions in $A{\rm  CrF}_3$, opening up the possibility
  of varying  the composition to  create novel states of  matter with
  tuneable properties.
\end{abstract}
\maketitle
\section{Introduction}
\label{sec:int}

\begin{figure*}[t!]
  \centering
 \includegraphics[scale=.6]{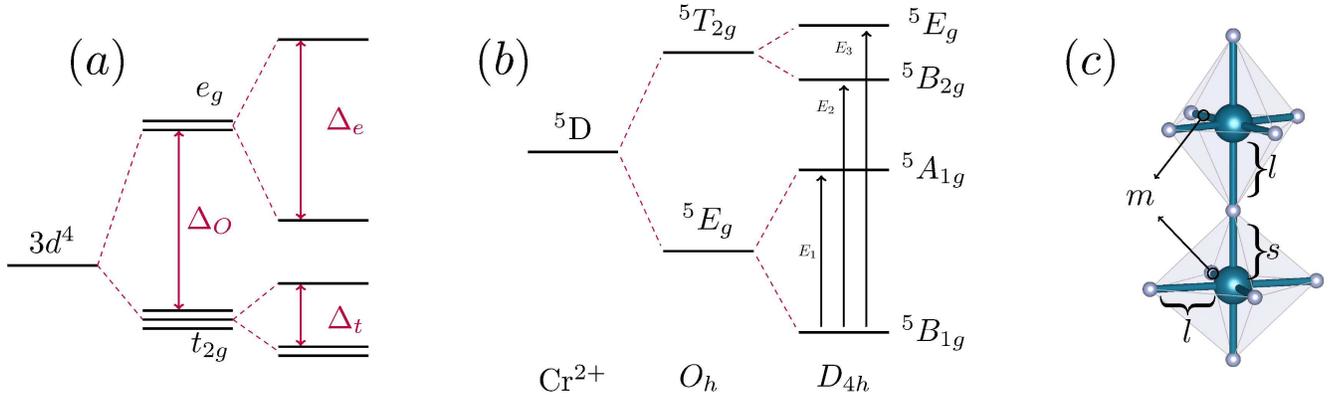}
 \caption{ ($a$) Evolution  of the energy level diagram  from the free
   $\mathrm{Cr^{2+}}$ ions into an octahedral crystal field $\Delta_O$
   followed  by  its  tetragonal  JT distortion  with  two  tetragonal
   splitting  parameters  $\Delta_e$  and $\Delta_t$,  and  its  three
   spin-allowed   transitions.   ($b$)   Energy-state  diagram   under
   symmetry  reduction  to  $D_{4h}$  owing to  the  JT-effect,  again
   showing the  three spin-allowed transitions $E_1$,  $E_2$and $E_3$.
   ($c$)  A pair  of  distorted octahedra  displaying the  short($s$),
   medium ($m$) and long ($l$) bonds.  }
   \label{fig:f0}
\end{figure*}


The  properties of  Jahn-Teller (JT)  active materials  arise from  an
intriguing  physical  interplay  between their  crystal,  orbital  and
magnetic structures \cite{COMP}.  The hallmark of JT-active systems is
orbital frustration (i.e.  orbital occupancy degeneracy).  This causes
structural distortions  that break the  symmetry and lift  the orbital
degeneracy \cite{JT}.

In  crystal field  theory,  the energy  levels of  an  ion with  $d^n$
electronic structure in  an octahedral field will split  into two sets
of energy levels:  threefold $t_{2g}$ and twofold  $e_g$, separated by
an octahedral  splitting energy  $\Delta_O$.  When  ions with  a $d^4$
electronic configuration such as ${\rm  Mn^{3+}}$ are at the center of
a  {[$BX_6^{n-3}$]} octahedron  (where $B$  is a  cation and  X is  an
anion) there is a further rearrangement of the energy levels, to lower
the energy  of the highest  occupied orbital, creating  two tetragonal
splitting energies: $\Delta_e$ and  $\Delta_t$.  This is the JT-effect
(See \fref{f0} ($a$)).   In a system transforming  from the octahedral
point  symmetry  $O_h$ to  JT  distorted  $D_{4h}$ (\fref{f0}  ($b$)),
vibrational  modes are  introduced corresponding  to axial  stretching
($Q_{\vartheta}$)  and   basal  squeezing   ($Q_{\varepsilon}$).   The
interaction    of    a     doubly    degenerate    electronic    state
$E_g(x^2-y^2,     3z^2-r^2)$    with     the    two     $e_g$    modes
$(Q_{\vartheta},  Q_{\varepsilon})$  is  known  as  the  $E\otimes  e$
problem (See Ref.  \cite{PRB60}).  Any point $\rho$ where:

\begin{equation}
  \label{eq:e0}
\rho=\sqrt{Q_{\vartheta}^2+Q_{\varepsilon}^2}\approx Q_{\vartheta}
\end{equation}

in the ($Q_{\vartheta},Q_{\varepsilon}$) space  will correspond to the
ground state  (since $Q_{\vartheta} >> Q_{\varepsilon}$).   The energy
gained  by having  a  singly occupied  rather  than doubly  degenerate
orbital state is known as the JT stabilization energy, $E_{JT}$, which
is proportional to $\rho$.

The  energy level  diagram  (\fref{f0} ($b$))  shows  the point  group
symmetry ($D_{4h}$) that develops under JT distortion. There are three
possible spin  allowed transitions from the  ${}^5B_{1g}$ ground state
of  the  $D_{4h}$  point   group:  $E_1  ({}^5B_{1g}\to  {}^5A_{1g})$,
$E_2  ({}^5B_{1g}\to {}^5B_{2g})$  and $E_3  ({}^5B_{1g}\to {}^5E_g)$.
These spin allowed transitions are defined by the tetragonal splitting
parameters and  the $E_{JT}$  and are  related through  the octahedral
perturbative   scheme   \cite{PRB60,   JCP}:

\begin{equation}
  \label{eq:eA}
  E_1=\Delta_e=4E_{JT}=K_e Q_{\vartheta}
\end{equation}

$E_2=\Delta_{(eq)}$  and  $\Delta_t=E_3-E_2=K_t Q_{\vartheta}$,  where
$K_e$ and  $K_t$ are the  electron-ion couplings.  It is  important to
note  that all  of  these  phenomena can  be  explained  by the  local
symmetry of a single octahedron.

Perovskites are  crystal structures with the  chemical formula $ABX_3$
based   on    a   network    of   vertex-sharing    octahedral   units
[$B^{n+}X_6^{m-n}$],  connected through  the {\em  B}-{\em X}-{\em  B}
angle ($\xi^\circ$), the octahedral tilt  or perovskite angle.  For an
ideal cubic perovskite  $\xi^\circ$ = 180$^\circ$. In  the presence of
JT-active ions, the distortions of the octahedra propagate through the
crystal lattice, causing $\xi^\circ$ to move away from 180$^\circ$ and
favoring the emergence of long  range orbital ordering (OO).  This has
consequences for  the magnetic structure of  the JT-active perovskite.
The ferromagnetic (FM)  interactions occur through antiferrodistortive
orbital ordering (AOO), e.g.  $\ket{3x^2-r^2}/\ket{3y^2-r^2}$.

X-ray diffraction and optical absorption (OA) spectroscopy experiments
under external  stimuli (e.g.  pressure, temperature,  magnetic field)
can be used to study the  interplay between the structure and internal
degrees of freedom in JT-compounds.  Among the most studied JT systems
are   the  ternary   alkali   metal/manganese  (III)   low-dimensional
fluoroperovskites   $A'_x\mathrm{MnF}_{3+x}$   (where   $A'$   is   an
alkali/alkali-earth  metal  and the  dimensionality  is  given by  the
number of vertex connections between the ${\rm MnF^{3-}_6}$ octahedra,
so  the $x=3,2,1$  are  0-,1- and  2-dimensional, respectively).   The
tetragonal   splitting  parameters   $\Delta_e$   and  $\Delta_t$   of
$A'_x\mathrm{MnF}_{3+x}$   increase   in    a   linear   manner   with
dimensionality \cite{JCP}.  This manifests itself in the OA spectra as
a   second   set   of   sharp   bands   representing   the   spin-flip
transitions. For this family of  materials the spin flip bands overlap
the spin  allowed bands $E_1$, $E_2$  and $E_3$.  Valliente et  al and
Rodriguez et al suggest that  oscillating intensities of the spin-flip
bands are correlated to the  perovskite angle $\xi^\circ$.  This angle
also controls  the symmetry  allowed overlap  between the  orbitals of
adjacent ions \cite{MF5} and  thus the magnetic exchange interactions.
The oscillating  strength of the  spin-flip bands also depends  on the
size of the ion at the $A$-site.  The ionic radii of the alkali metals
(group 1  of the  periodic table)  increase significantly  with atomic
number, providing a good test case for this.

The  2-dimensional   fluoride  $\mathrm{CsMnF_4}$  is   a  transparent
ferromagnet which  adopts the  tetragonal space  group $P4/n$  at room
temperature   \cite{MF1}.    The  magneto-structural   properties   of
$\mathrm{CsMnF_4}$ at high pressure have been thoroughly studied. High
pressure    synchrotron   XRD    (SXRD)   experiments    showed   that
$\mathrm{CsMnF_4}$  goes  through a  tetragonal-to-orthorhombic  phase
transition  from $Pa/n\to  Pmab$  at about  $\sim$1.6 GPa  \cite{MF3}.
Magnetic susceptibility measurements show  that this is accompanied by
a ferromagnetic to antiferromagnetic (FM-to-AFM) phase transition.  OA
spectroscopy  reveals a  high-to-low spin  transition at  37.5 GPa  in
which  the   three  spin-allowed  transition  bands   merge  into  one
\cite{MF5}.   The  explanation  for  this  is  that  as  the  pressure
increases  $E_{JT}$  is  surpassed  by the  octahedral  crystal  field
splitting   energy,  $\Delta_O$,   forcing  ${\rm   Mn^{3+}}$into  the
octahedral low  spin configuration.  This high-to-low  spin transition
is not observed for the  smaller $A'$ -site atoms, $\mathrm{Na^+}$ and
$\mathrm{K^+}$\cite{MF2,MF3,MF4}.

Chromium (II) fluoroperovskites have been  difficult to study owing to
the   oxygen-sensitive    chemistry   of    $\mathrm{Cr^{2+}}$   ions.
$\mathrm{KCrF_3}$ (which can  be prepared by a solid  state route) has
three  temperature   dependent  polymorphs:   cubic  ($T>$   1000  K),
tetragonal ($T$ = 1000 to 250 K) and monoclinic ($T<250$ K); where the
tetragonal to cubic  phase transition is suggested to  correspond to a
metal-insulator transition  (MIT) as  the OO  is removed  and orbitals
become degenerate\cite{SM1,SM2}.  There are in addition three magnetic
phases (incommensurate-AFM  (IC-AFM) between  300 and 79.5  K ($T_N$),
commensurate-AFM (C-AFM)  between $T_N$ and  45 K, weak FM  and canted
AFM below 10 K) \cite{PRB82}.

We  recently developed  a synthesis  route for  $\mathrm{NaCrF_3}$ and
were able  to explore its  structural and magnetic properties  for the
first time \cite{ME}. It has a triclinic structure ($P\bar{1}$) at 300
K with  a single  canted-AFM magnetic  structure below  21 K\cite{ME}.
Measurements of field  dependent magnetization - ($ M(H)  $) below the
N\'eel  temperature   have  revealed  the  onset   of  a  metamagnetic
transition at an applied field of 8 T.

In this paper we study the importance  of the $A$-site ion size in the
three  dimensional perovskite  family  $A\mathrm{CrF_3}$  where $A$  =
Na$^+$ (small) or K$^+$  (large) using temperature and field-dependent
OA  spectroscopy,  supported by  crystal  structure  analysis and  DFT
calculations.  In general we see that as the symmetry of the system is
reduced the long  range magnetic interactions become  weaker and local
magnetic structures begin to dominate the behaviour.
Our results show that the changes  in the spin-flip transitions at low
temperature are not correlated to the perovskite angle $\xi^\circ$, but
rather  to  the  magnetic exchange  interactions  between  neighboring
${\rm CrF_6^{4-}}$ octahedra. 
Merging  of  the spin-allowed  spectral  bands  $E_2$ and  $E_3$  for
$A  =  {\rm Na^+}$  shows  the  emergence  at  low temperature  of  an
intermediate electronic state where $\Delta_t$ is zero.
Furthermore, when an increasing external  magnetic field is applied at
2  K,  the   intensity  of  the  spin  allowed   bands  increases  for
${\rm NaCrF_3}$ but not for ${\rm KCrF_3}$. 

$E_{JT}$  is larger  in ${\rm  NaCrF_3}$ compared  to ${\rm  KCrF_3}$,
owing to the smaller size of the $Na^+$ ion. Again, we believe this is
explained by the  local orbital and magnetic ordering  rather than the
perovskite angle $\xi^\circ$.
We  also present  density  functional theory  (DFT) simulations  which
calculate  the superexchange  parameters  for  the different  magnetic
orderings possible in each crystal structure, and so, determine the OO
of each phase. Finally, we show that the synthesis method we described
for  ${\rm NaCrF_3}$  \cite{ME}  can  also be  used  to produce  large
volumes of high purity ${\rm KCrF_3}$.

\section{Experimental Methods}
\label{sec:meth}

${\rm  NaCrF_3}$ was  prepared according  to the  method described  by
Bernal  et al.   \cite{ME} ${\rm  KCrF_3}$ was  prepared by  a similar
method:    0.5     g    of     chromium    (II)     acetate    hydrate
(${\rm  Cr_2(CH_3CO_2)_4  (H_2O)_2}$; CrOAc)  was  dissolved  in 2  ml
deagassed water in a polycarbonate vial sealed with a silicone rubbber
septum under  constant flow of  Ar.  In a second  sealed polycarbonate
vial 0.3 g KHF$_2$ was dissolved  in 10 mL degassed water. The KHF$_2$
solution was then injected with a dry syringe into the CrOAc solution.
${\rm KCrF_3}$  precipitated immediately.  Washing with  pure methanol
rendered both samples air stable.  The washed samples were dried under
vacuum overnight, and then stored in a glove-box.

Powder X-ray diffraction (XRD) measurements for $\mathrm{KCrF_3}$ were
performed  at the  Norwegian  Resource Centre  for X-ray  Diffraction,
Scattering and Imaging (RECX) on a Bruker D8 Advance diffractometer in
capillary mode with $\mathrm{Cu_{k\alpha}}$ radiation selected by a Cu
(111) focusing  monochromator and a LynxEye  XE detector.  Synchrotron
powder  XRD   data  for  $\mathrm{NaCrF_3}$  were   collected  at  the
Swiss-Norwegian  Beam line  (SNBL) BM01A  of the  European Synchrotron
(ESRF),   Grenoble,  France.    The  setup   is  described   elsewhere
\cite{SNBL}.  The diffuse reflectance  spectra (DRS) were collected on
polycrystalline  samples of  $\mathrm{NaCrF_3}$ and  $\mathrm{KCrF_3}$
with a  UV-3600 photospectrometer (Shimadzu).  The  low temperature OA
experiments were  conducted under  vacuum in  a closed  cycle cryostat
(Janis Research)  with a Fibre-Lite MI-150  halogen lamp (Dolan-Jenner
Industries) and USB4000 spectrometer  (OceanOptics) in the temperature
range 11 - 300 K, and from 1.3 to 2.7 eV.

Field dependent magnetization measurements were  carried out on the AC
magnetic  measurement system  -II (ACMS-II)  using a  vibrating sample
magnetometer  (VSM)  on  the  Quantum Design  (QD)  Physical  Property
measurement system (PPMS).   Samples of known weight  were loaded into
weakly  diamagnetic plastic  holders and  immobilized on  brass sample
holders at  defined height of approximately  25 mm from the  bottom of
the brass holder.  This holder was then loaded into the ACMS-II at the
end of a long  plastic rod and inserted into the  PPMS and cooled down
to  2 K.   Sample was  kept  at 2  K  for 30  minutes before  starting
magnetization   measurements  for   thermal   homogenization  of   the
sample. The magnetization measurements were carried in sweep mode from
0 to  +/- 9 T at  a ramp rate of  0.01 T/second and a  measurement was
taken each second.

Magneto-optic measurements were  carried out in a  Quantum Design PPMS
system.   A custom  insert  was developed  using  an external  DH-2000
Halogen lamp (Mikropack) and  USB4000 spectrometer (OceanOptics), both
optically connected  with the insert  using an optical fibre.   A long
silica rod with the sample powder attached in a gelatin capsule at one
end was used  to bring light in  and out of the  vacuum sample chamber
and to  position the sample  appropriately with respect to  the PPMS's
superconducting magnets.   A LakeShore Cryotronics  CERNOX thermometer
was utilized for reading local sample temperature close to the sample.

For temperature dependent measurements,  the silica rod mounted sample
was inserted into  the PPMS cavity and the entire  space was evacuated
to a  soft vacuum  at room  temperature.  The  sample and  cavity were
thermally stabilized before the cool down process was initiated.  Cool
down from 300 to 2 K was carried out  at the rate of 0.5 K/ min and an
optical spectrum was taken every  10 seconds.  Once the system reached
the 2 K temperature set-point a 30 minute delay was given in order for
the  sample  temperature  to  stabilize, all  the  while  continuously
measuring the  OA spectra.  Magnetic field  dependent OA measurements,
were made at 2 K and the magnetic field was ramped at 0.01 T/second up
to 9 T, continuously measuring OA.

The  electronic  structures   of  the  $\mathrm{NaCrF_3}$  (triclinic-
$\mathcal{I}$)   and  $\mathrm{KCrF_3}$   (monoclinic-  $\mathcal{M}$,
tetragonal-  $\mathcal{T}$,  and  cubic-  $\mathcal{C}$)  phases  were
calculated  using  the  Vienna  \emph{ab  initio}  simulation  package
(VASP)\cite{VA1,VA2} with the PBE general gradient approximation (GGA)
\cite{PBE}.    $\mathrm{KCrF_3}$-  $\mathcal{C}$   was  included   for
comparison, as  the highest possible  symmetry.  The cutoff  energy of
the plane wave  basis set expansion was at least  450 eV.  The density
of the $k$ point density was  determined by a maximum distance between
points  of 0.25  \AA$^{-1}$.   Hubbard-corrected GGA+$U$  calculations
were performed according  to the method described  by Meredig \emph{et
  al.}  \cite{DFT005} with the Coulomb interaction parameter $U$ going
up to $U=9$  eV and the exchange interaction  parameter $J_H=0.88$ for
chromium 2+.  For the calculation  of the superexchange parameters, we
used  the same  parameters  as  in the  GGA+$U$  calculations on  four
possible configurations:

\begin{itemize}
\item \textbf{F}, (ferromagnetic) in which all the spins are parallel;
\item \textbf{A},  (antiferromagnetic) in  which the  intralayer spins
  are parallel while the interlayer spins are antiparallel;
\item \textbf{C}, in which the chains  of spins along the $z$-axis are
  ferromagnetically  coupled,   while  adjacent  chains   are  aligned
  antiparallel;
\item   \textbf{G},  in   which   all  nearest   neighbor  spins   are
  antiparallel.
\end{itemize}

\subsection{Structural Analysis of  $A\mathrm{CrF_3}$}

\begin{figure*}[t!]
  \centering
 \includegraphics[scale=.72]{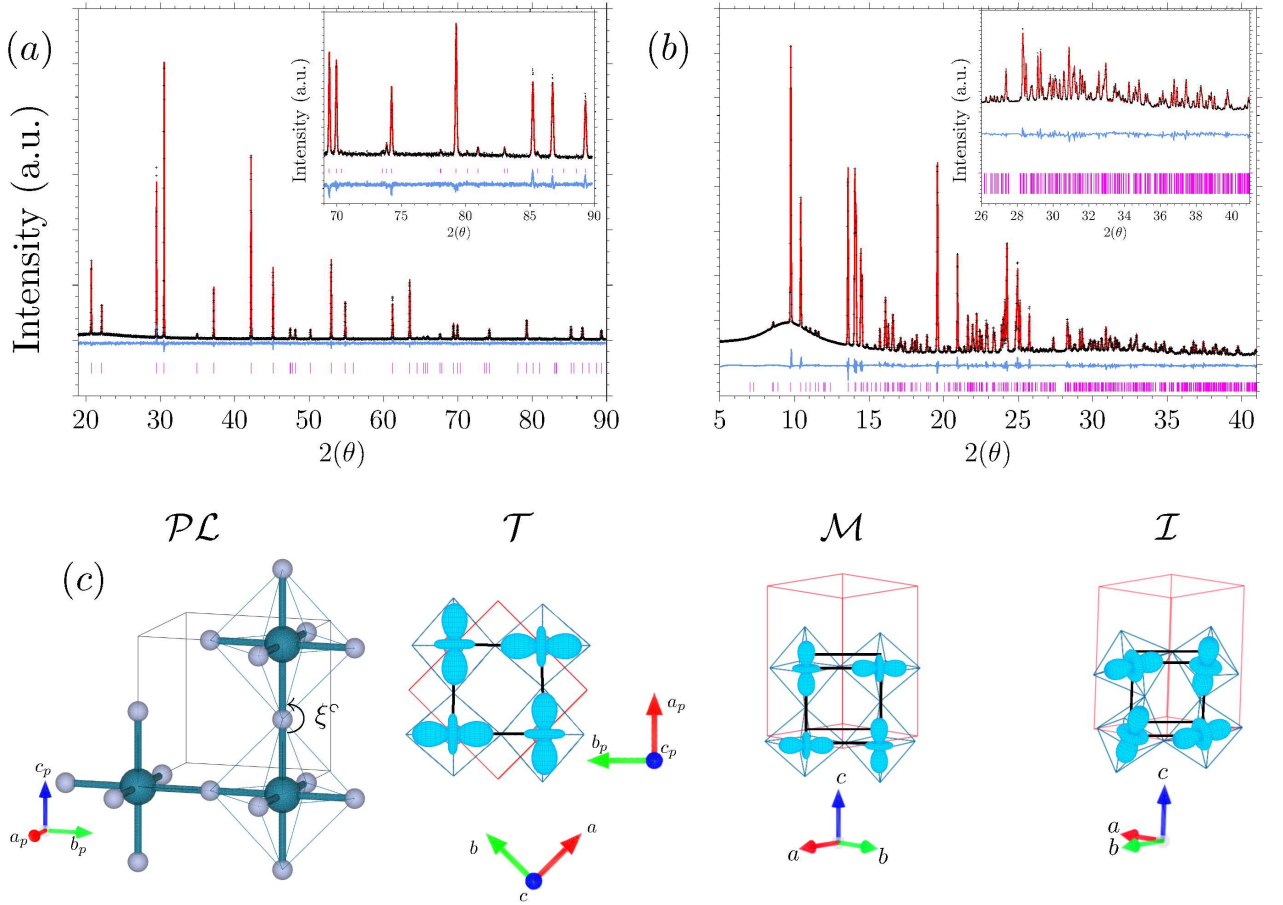}
 \caption{($a$) Observed (black dots) and calculated (red
   line) 
   X-ray    powder   diffraction    (XRD)   for    T-$\mathrm{KCrF_3}$
   ($\lambda=1.5405$ \AA)  at 258 K ($a=6.051317$,  $c=8.024635$ \AA).
   Agreement factors of the Rietveld refinement: $R_{wp}=16.09476$ \%;
   $R_{exp}=5.3272$  \%.  The  solid  blue line  shows the  difference
   between  the observed  and  calculated intensities,  and the  ticks
   indicate the  allowed Bragg reflection positions.   The inset shows
   the high angle fit.  ($b$)  Final observed Synchrotron XRD profiles
   ($\lambda=0.709$ \AA\  of $\mathrm{NaCrF_3}$ at 298  K ($a=5.5100$,
   $b=5.69190$, and  $c=8.18860$ \AA).  The agreement  of the Rietveld
   refinement: $R_{wp}=3.321219$ and  $R_{exp}=0.014351$ \%.  the very
   low $R_{wp}$ value is due to the use of an area detector (ref).
   ($c$)  Schematic  representations  of,  from  left  to  right:  the
   pseudocubic unit  cell with $a_p=b_p=c_p$ ($\mathcal{PL}$),  the PL
   embedded   within   the  tetragonal   ($\mathcal{T}$),   monoclinic
   ($\mathcal{M}$)  and   triclinic  ($\mathcal{I}$)  unit   cells  of
   $A{\rm CrF_3}$.   The OO  motif of the  $\ket{d_{i^2}}$ ($i=x,y,z$)
   orbitals is shown.}
   \label{fig:f1}
\end{figure*}

\begin{table}[ht!]
  \centering 
  \caption{  Octahedral  tilt  angles  $\xi^\circ  (\mathcal{I})$  and
    distortions   ($\Delta   d$)   from  XRD   for   $A{\rm   CrF_3}$.
    $\xi^\circ     (\mathcal{I})$    for     $\mathrm{NaCrF_3}$    and
    $\xi^\circ (\mathcal{M})$  for $\mathcal{M}$-$\mathrm{KCrF_3}$ are
    taken  from \cite{PRB82}.  $\xi^\circ (\mathcal{T})$  is fixed  to
    $180 ^\circ$ ($\xi^{\pi}$) by symmetry.  The calculated octahedral
    distortions  $\Delta  d$  ($\times   10^{-4}$)are  given  for  all
    phases.}
  \label{tab:T1} 
  \begin{tabular}{l c c c c c c c c c c c c c}
    \hline\hline
    &&&&&&&&&&&&&\\    
    Cr&&$\Delta d^{\mathrm{\mathcal{T}}}$&& $\Delta d^{\mathrm{\mathcal{M}}}$&&$\Delta d^{\mathrm{\mathcal{I}}}$&&  && &&&  \\
    &&&&&&&&&&&&&\\
    1:&&65.70 &&56.53 &&78.85 & &&&&&\\
    &&&&&&&&&&&&&\\
    2:&&-&&69.97 &&55.86 &&&&&&\\
    &&&&&&&&&&&&&\\
    3:&&-&&-&&76.24& &&&&\\
    &&&&&&&&&&&&&\\
    4:&&-&&-&&70.73 && &&&&\\
    &&&&&&&&&&&&&\\
    &&&& $\xi^\circ$($\mathcal{M}$)  &&&&\\
    &&&&&&&&&&&&&\\
    Cr1-F3-Cr2 &:&      &&         &&       180.0 &&&	$l$-axial&&&&\\
    &&&&&&&&&&&&&\\
    Cr1-F2-Cr2 &:&      &&         &&       167.3&&&	$sm$-planar\\
    &&&&&&&&&&&&&\\
    Cr1-F1-Cr2 &:&       &&        &&       162.3&&&	$sm$-planar\\
    &&&&&&&&&&&&&\\
    &&&& $\xi^\circ$($\mathcal{I}$)  &&&&\\
    &&&&&&&&&&&&&\\
    Cr1-F5-Cr3 &:&        &&       &&       138.797 &&&	$l$-axial&&&&\\
    &&&&&&&&&&&&&\\
    Cr2-F2-Cr1 &:&        &&        &&       139.446 &&&	$sm$-planar\\
    &&&&&&&&&&&&&\\
    Cr2-F1-Cr1 &:&        &&        &&       146.589 &&&	$sm$-planar\\
    &&&&&&&&&&&&&\\
    Cr4-F6-Cr2 &:&        &&        &&        140.526 &&&	$l$-axial\\
    &&&&&&&&&&&&&\\
    Cr4-F4-Cr3 &:&        &&        &&        142.357 &&&	$sm$-planar\\
    &&&&&&&&&&&&&\\
    Cr4-F3-Cr3 &:&        &&        &&        141.683 &&&  $sm$-planar\\
    &&&&&&&&&&&&&\\
    \hline
  \end{tabular}
\end{table}

To analyze the $A\mathrm{CrF_3}$ systems, we use a pseudocubic lattice
(PL) derived from the unit cell  of the high T cubic $\mathrm{KCrF_3}$
phase.    Its  vertices   lie  at   the  centers   of  8   neighboring
$\mathrm{CrF_6^{4-}}$  octahedra,  and   its  lattice  parameters  are
$a_p{\approx}b_p{\approx} c_p$. The JT  distorted octahedra have short
($s$, ca.  1.9 $\AA$),  medium ($m$,  ca.  2.0  $\AA$) and  long ($l$,
ca. 2.3 $\AA$)  Cr-F bonds, the relative orientation  of which depends
on the  orbital ordering. The  $s$, $m$ and $l$  bonds can be  used to
describe the orientations of isolated $\mathrm{CrF_6^{4-}}$ octahedra.
For example, the  tilt angle $\xi^\circ_l$ along the  $l$-bonds can be
described as $l$-axial, and if a  bond lies within the plane formed by
the $s$ and $m$ bonds, then, it  is described as the $sm$-plane of the
octahedra.  The octahedral  distortion  parameter  $\Delta_d$ used  to
quantify the distortion (ref Shannon) is also calculated from the $s$,
$m$ and $l$ Cr-F bond lengths around the octahedron:

\begin{equation}
\label{eq:e1}
\Delta_d=1/6\sum_{i=1}^6  |{\rm l}_i-{\rm l}_{av}|/{\rm l}_{av}
\end{equation}

Where l$_i$ for  i = 1-6 is the  6 Cr-F bond lengths, (2  each of $s$,
$m$  and  $l$)  and l$_{av}$  is  the  average  of  $s$, $m$  and  $l$
\cite{BAUR}.

In general we  can describe all of the structures  as having layers in
the $s$, $l$ plane where  neighboring octahedra are rotated 90$^\circ$
about the $m$-axis  to create a motif of alternating  $s$ and $l$ Cr-F
bonds. These  layers are linked by  $m$ Cr-F bonds. The  PL is aligned
relative to the tetragonal ($\mathcal{T}$), monoclinic ($\mathcal{M}$)
and triclinic  ($\mathcal{I}$) crystallographic  unit cells  such that
the network  of alternating $s$ and  $l$ bonds lies in  the $a_p, b_p$
plane and the linking $m$ bonds are aligned along $c_p$ (see \fref{f1}
(c)).   \fref{f1}  ($c$) shows  the  relationship  of  the PL  to  the
($\mathcal{T}$), ($\mathcal{M}$) and ($\mathcal{I}$) unit cells of the
$A{\rm  CrF_3}$  family.  The  tilts  ($\xi^\circ$)  of the  octahedra
relative to  the edges  of the  ideal PL increase  as the  symmetry is
lowered. The layers of alternating $s$ and $l$ bonds in the $a_p, b_p$
plane are  stacked with adjacent  layers mirrored in the  $a_p$, $c_p$
plane. This  creates an A, B,  A, B... sequence. At  the orbital level
this       means       that       the       OO       consists       of
$\ket{d_{3x^2-r^2}}/\ket{d_{3y^2-r^2}}$ AOO  layers stacked  along the
$c_p$-axis.

In an  ideal cubic perovskite all  $\xi^\circ=180$ ($\xi^{\pi}$).  The
higher  the  number of  $\xi^{\pi}$  vertex-angles  in a  system,  the
stronger  the magnetic  exchange  interactions due  to proper  orbital
overlap.  To  the ideal  system described  by the  PL, we  applied the
perturbative  model  \cite{PRB60,  JCP}  to  describe  the  structural
parameters  related  to  the  JT-distortions  from  the  XRD-patterns.
Inspection of  the powder XRD  pattern (\fref{f0}) confirmed  that our
syntheses       yielded       single       phase       polycrystalline
$\mathcal{T}$-$\mathrm{KCrF_3}$    (space    group    $I4/mcm$)    and
$\mathcal{I}$-$\mathrm{NaCrF_3}$  (space  group  $P\bar{1}$)  at  room
temperature. Octahedral distortion parameters $\Delta d$ for the three
$A\mathrm{CrF_3}$ phases, and perovskite tilt angles $\xi^{\circ}$ for
for $\mathcal{M}$-${\rm KCrF_3}$and  $\mathcal{I}$-${\rm NaCrF_3}$ are
give in \tref{T1}.

$\Delta d$  for $\mathcal{T}$-${\rm KCrF_3}$ is  65.7$\times 10^{-4}$.
It  adopts the  zero-tilt configuration  described by  Glazer notation
$a^\circ a^\circ a^\circ$ \cite{GLAZER} so that all periovskite angles
are $\xi^{\pi}$.  The $\xi^{\pi}$ angles  within the AOO layers create
strong in-plane FM interactions propagating  along the $a_p$ and $b_p$
axes.   The magnetic  superexchange parameters  within the  AOO-planes
display FM character through $\sigma$ interactions between half-filled
$\ket{d_{3z^2-r^2}}$ and empty  $\ket{d_{x^2-y^2}}$ Cr$^{2+}$ orbitals
bridged by $\ket{p_{\sigma}}$ orbitals  of F$^-$ ions.  The orthogonal
$\xi^{\pi}$ perovskite angle  along the $c_p$-axis (absent  in the low
dimensional    $A'_x\mathrm{MnF}_{3+x}$    systems)   leads    to    a
ferrodistortive  OO  (FOO)  with  an  AFM  coupling  between  adjacent
overlapping empty $\ket{d_{x^2-z^2}}$ and $\ket{d_{y^2-z^2}}$ orbitals
bridged by F$^- \ket{p_{\sigma}}$ orbitals.

$\mathcal{M}$-$\mathrm{KCrF_3}$  (space group  $  I112/m$) appears  on
cooling   to   $T=$250   K   \cite{SM1}.    Two   crystallographically
non-equivalent Cr  sites (Cr1,  Cr2) are generated  in this  form with
$\Delta d$ values of  56.53$\times 10^{-4}$ and 69.97$\times 10^{-4}$.
Within  the  $a_p,  b_p$  and   $b_p,  c_p$  planes  there  are  tilts
corresponding to Glazer notation $a^-b^-c^0$, in which, the Cr1-F1-Cr2
and Cr1-F2-Cr2 angles are 162 and 167$^\circ$, respectively.  The loss
of  symmetry  leaves  a  single $\xi^{\pi}$  vertex  angle  along  the
$c_p$-axis  where   proper  orbital   overlap  is  preserved   for  FM
interactions to  propagate.  The orbital overlap  for AFM interactions
also becomes weaker  which can be interpreted as  weak FM interactions
since      the      $\ket{d_{x^2-z^2}}-\ket{p_{\sigma}}-\ket{y^2-z^2}$
relationship is weakened.

$\mathrm{NaCrF_3}$ crystallizes in a $\mathrm{NaCuF_3}$ type structure
with   space  group   $P\bar{1}$   \cite{NCUF}.   Four   nonequivalent
crystallographic sites (i.e.   Cr1, Cr2, Cr3 and Cr4)  are present for
$\mathrm{Cr^{2+}}$  in $\mathrm{NaCrF_3}$  \cite{ME}, with  $\Delta d$
and $\xi^{\circ}$ values  give in table 1.  The $\Delta  d$ values for
$\mathrm{NaCrF_3}$  are greater  than those  of the  $\mathrm{KCrF_3}$
phases, indicating  greater JT distortion.   The $\mathrm{CrF_6^{4-}}$
units  are  tilted  in  all   directions,  adopting  the  Glazer  tilt
$a^-b^-c^-$.   This  means that  ${\rm  NaCrF_3}$  has no  $\xi^{\pi}$
angles, which is reflected in  a significant reduction in the strength
of the overall magnetic interactions.   Consequently, the AOO adopts a
canted motif where $\ket{d_{z^2}}$ orbitals  on Cr1 and Cr4 tilt about
the $b_p$-axis and Cr2 and Cr3 are tilted about the $a_p$-axis.

\begin{figure}[t!]
  \centering
  \includegraphics[scale=.65]{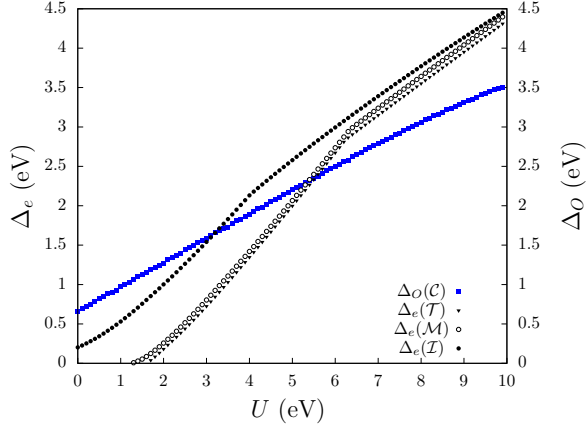}
  \caption{ Calculated octahedral $\Delta_O$ and tetragonal $\Delta_e$
    splitting parameters as function of $U$.  $\Delta_O (\mathcal{C})$
    octahedral  splitting of  the  cubic  phase of  $\mathrm{KCrF_3}$,
    tetragonal  $\Delta_e  (\mathcal{T})$  for the  tetragonal  phase,
    $\Delta_e   (\mathcal{M})$   for    the   monoclinic   phase   and
    $\Delta_e  (\mathcal{I})$  corresponding  to the  triclinic  space
    group in $\mathrm{NaCrF_3}$. }
  \label{fig:delta}
\end{figure}

\begin{figure}[t!]
  \raggedleft
  \includegraphics[scale=.7]{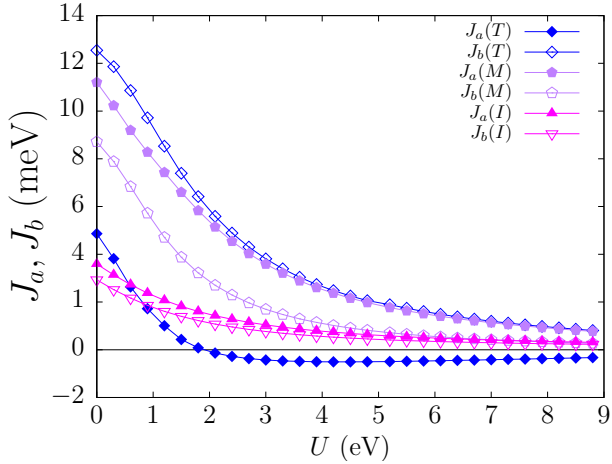}
  \caption{ Calculated  axial $J_A$ and basal  $J_B$ magnetic exchange
    parameters    for     $\mathcal{T}$-$\mathrm{KCrF_3}    (I4/mcm)$,
    $\mathcal{M}$-$\mathrm{KCrF_3}            (I112/m)$            and
    $\mathcal{I}$-$\mathrm{NaCrF_3} (P\bar{1})$-phases  as function of
    $U$.  Ferromagnetic exchange for $J_i>0$ and antiferromagnetic for
    $J_i<0$.}
\label{fig:f4}
\end{figure}

\subsection{Electronic Structure of $A\mathrm{CrF_3}$}
\label{sec:DFT}

The calculated octahedral and  tetragonal splitting parameters for the
$A\mathrm{CrF_3}$  phases,  as  a function  of  the  electron-electron
correlation  parameter  $U$,  are   presented  in  \fref{delta}.   For
$\mathcal{C}$-$\mathrm{KCrF_3}$  there is  no  JT  distortion and  the
octahedral splitting parameter $\Delta_O$ is  present at all levels of
correlations, preserving the orbital degeneracy of the $e_g$ orbitals.

This invariance under $U$ suggests that electron-electron correlations
lead to neither OO nor  MIT in $\mathcal{C}$-${\rm KCrF_3}$ within the
framework    of   our    calculations.     For   $\mathcal{T}$-    and
$\mathcal{M}$-$\mathrm{KCrF_3}$   phases,   the   emergence   of   the
tetragonal  splitting parameter  $\Delta_e$  occurs simultaneously  at
$ U{\approx}1.7$  eV.  This may  suggest that the  OO and MIT  in both
$\mathcal{T}$-  and  $\mathcal{M}$-${\rm  KCrF_3}$ phases  emerges  as
consequence   of   strong   electron   correlations.    However,   the
calculations  were performed  on  the already  distorted low  symmetry
${\rm KCrF_3}$  which excludes  electron-electron correlations  as the
source of the OO.  In  other words, structural distortion is necessary
for OO  to occur.   Replacing K$^+$  with Na$^+$  in $A\mathrm{CrF_3}$
increases $E_{JT}=\Delta_e/4$  even at $U=0$  and from this  point its
increase  with  $U$ has  a  similar  curvature to  $\mathcal{T}$-  and
$\mathcal{M}$-$\mathrm{KCrF_3}$.    Our   calculations   support   the
structural results where $E_{JT}$, which is related to $Q_{\vartheta}$
(Equation   \ref{eq:eA}),  is   larger   for   ${\rm  NaCrF_3}$   than
${\rm  KCrF_3}$.  At  $U=0$  ${\rm  NaCrF_3}$  preserves  the  orbital
degeneracy  of the  $e_g$  orbitals showing  that the  phonon-electron
interactions that characterise the JT-effect are the driving force for
OO.

\begin{figure}[t!]
  \flushleft
 \includegraphics[scale=.65]{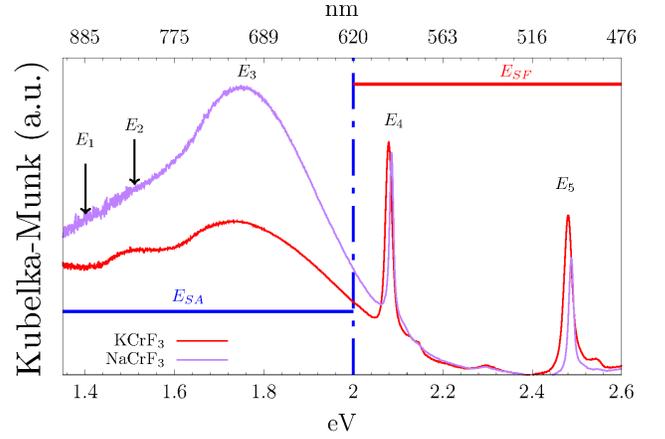}
 \caption{ OA  spectra of $\mathrm{KCrF_3}$ and  $\mathrm{NaCrF_3}$ at
   300 K.  Spin-allowed (quintet-quintet) $E_{SA}$ transitions: $E_1$,
   $E_2$ and $E_3$.  Spin-flip (quintet-triplet) $E_{SF}$ transitions:
   $E_4$  and  $E_5$.}  
 \label{fig:f8}
\end{figure}

\begin{figure*}[t!]
  \centering
 \includegraphics[scale=.65]{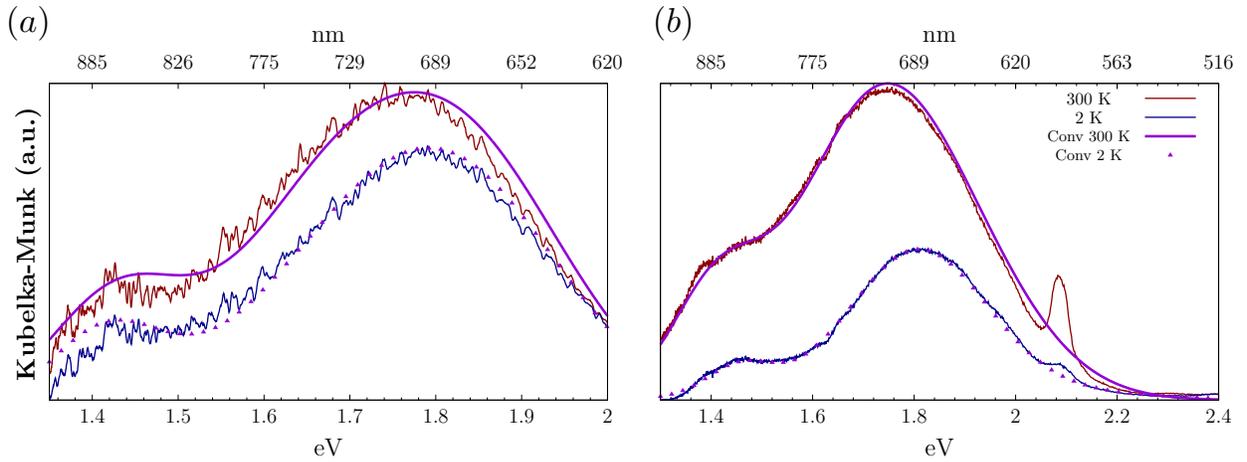}
 \caption{OA   spectra   of    ($a$)   $\mathrm{KCrF_3}$   and   ($b$)
   $\mathrm{NaCrF_3}$ at  300 and  2 K  corresponding to  the $E_{SA}$
   region with their respective convolution fits. }
 \label{fig:f2}
\end{figure*}

\begin{figure}[t!]
  \flushleft
 \includegraphics[scale=.71]{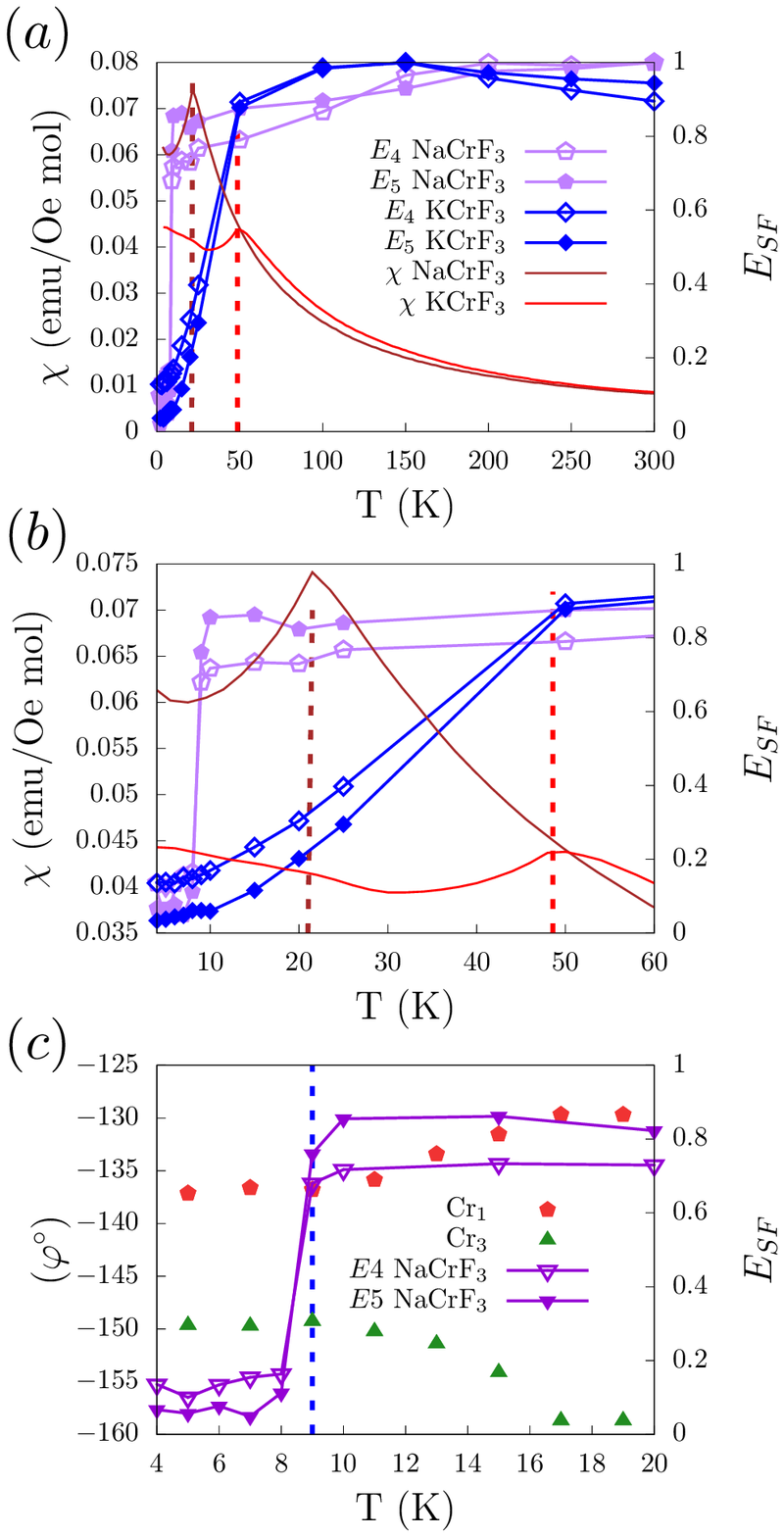}
 \caption{($a$) Temperature dependent OA of $E_{SF}$ transitions $E_4$
   and  $E_5$ for  $\mathrm{KCrF_3}$  and  $\mathrm{NaCrF_3}$ and  ZFC
   magnetic susceptibility $\chi$ in the interval 2 to 300 K.  The two
   vertical  dashed lines  at $T_N=21.5$  K (brown)  and $T_1=48.6$  K
   (red) indicate the Ne\'el temperature of $\mathrm{NaCrF_3}$ and the
   IC-to-AFM  transition  in $\mathrm{KCrF_3}$,  respectively.   ($b$)
   Close-up of the $E_{SF}$ and $\chi$  in the range of 4-60 K.  ($c$)
   Polar  angle  $\vartheta$  plotted   with  $E_{SF}$  intensity  for
   ${\rm NaCrF_3}$.  The vertical blue  dashed line  at 9 K  marks the
   reduction of oscillating strength and  the final convergence of the
   polar angle for Cr1 and Cr3 Ref.\cite{ME}}
 \label{fig:f7}
\end{figure}

We   calculated  the   ground-state  energies   of  the   low-symmetry
$A{\rm CrF_3}$ phases with four magnetic orderings ${\bf A,F,C,G}$ and
then  the  magnetic exchange  parameters  $J_{B}$  (basal) within  the
$a_pb_p$-plane of the pseudocubic unit cell, and $J_{A}$ (axial) along
the               $c_p$-axis               according               to:
$J_{A}=[E ({\bf F})  -E ({\bf G})- E ({\bf A})+E  ({\bf C})]/4S^2$ and
$J_{B}=[E  ({\bf F})  -E ({\bf  G})+  E ({\bf  A})-E ({\bf  C})]/8S^2$
\cite{PRB84}.   \fref{f4}  shows  the energies  of  the  superexchange
parameters  as function  of  $U$ for  the  tetragonal, monoclinic  and
triclinic phases of $A{\rm CrF_3}$.
For  $\mathcal{T}$-${\rm KCrF_3}$  the calculated  axial superexchange
parameter is $J_{A}  (\mathcal{T}) >0$ for $U<1.9$,  and become weakly
negative for  $U\geq 2$ suggesting  weak AFM interactions  between AOO
layers.  The  basal superexchange  $J_{B} (\mathcal{T})$  decreases at
all   levels   of   correlations   but   is   always   positive   with
$J_{B} (\mathcal{T})>J_{A} (\mathcal{T})$.
This picture agrees with the  one described in the structural section,
in which $\mathcal{T}$-${\rm KCrF_3}$ contains only $\xi^{\pi}$ vertex
angles, both within its AOO  $a_p, b_p$-plane (with strong in-plane FM
interactions)     and    connecting     the     AOO    layers.      In
$\mathcal{M}$-${\rm  KCrF_3}$, the  loss of  symmetry leaves  a single
$\xi^{\pi}$  along  the   $c_p$-axis.   The  calculated  superexchange
parameters    of    $\mathcal{M}$-${\rm    KCrF_3}$    confirm    that
$J_{A}        (\mathcal{M}),J_{B}         (\mathcal{M})>0$        with
$J_{A}  (\mathcal{M})>J_{B}   (\mathcal{M})$,  the  opposite   of  the
tetragonal  form.  Again,  this agrees  with the  structural analysis,
describing a  single $\xi^{\pi}$  in the $c_p$-axis  direction, giving
strong FM interactions propagated along the $c_p$-axis.

In ${\rm  NaCrF_3}$, the  smaller Na$^+$ ion  on the  $A$-site induces
additional  octahedral tilts  in  all directions  with no  $\xi^{\pi}$
vertex   angles,  accompanied   with  larger   octahedral  distortions
(\tref{T1}).    This  further   reduces  strength   of  the   magnetic
interactions  as  the  orbital  overlap  decreases.   The  canted  AOO
propagates   within   the   $a_pc_p$-plane.    ${\rm   NaCrF_3}$   has
$J_{A} (\mathcal{I})\sim  J_{B} (\mathcal{I})>0$ with  increasing $U$.
Comparing  to the  higher symmetry  systems found  in ${\rm  KCrF_3}$,
$J_{A  (B)} (\mathcal{M})>J_{A  (B)}(\mathcal{I})$ indicates  that the
magnetic interactions are weakest at the lowest symmetry.
We                             observe                            that
$J_A  (\mathcal{T})>J_{A,B}  (\mathcal{M})>J_{A,B} (\mathcal{I})$  for
$U<4$ but  for $U\geq 4$  $J_A (\mathcal{T})$ and  $J_A (\mathcal{M})$
converge. Again this is consistent with the structural analysis, where
increasing distortion reduces orbital overlap.


\begin{table}[ht!]
  \centering 
  \caption{ Calculated  JT vibrational  modes and  fitted spin-allowed
    band positions in the OA spectra of $\mathrm{ACrF_3}$.  The normal
    modes are  calculated from  the bond  distances in  the octahedral
    units in \AA\ according to Equations 
    \ref{eq:e2}.  Fitted positions of  the $E^{300K}_{SA}$ bands in eV
    at    300   K.     Tetragonal   parameters    $E_1=\Delta_e$   and
    $\Delta_t=E_3-E_2$ along with the JT-stabilization energy $E_{JT}$
    given in eV. }
    \label{tab:T2} 
  \begin{tabular}{l c c c c c c c c c c c c c}
    \hline\hline
    &&&&&&&&&&&&&\\
    Cr&&$Q_{\vartheta}$($\mathcal{T}$)&&$Q_{\vartheta}$($\mathcal{M}$)&&$Q_{\vartheta}$($\mathcal{I}$)& $Q_{\varepsilon}(\mathcal{T})$&$Q_{\varepsilon}(\mathcal{M})$&& $Q_{\varepsilon}(\mathcal{I})$&& \\
    &&&&&&&&&&&&&\\
    1:&&0.33&&0.30&&0.43&0.019 & 0.071&& 0.042&&\\
    &&&&&&&&&&&&&\\
    2:&&-&&0.38&&0.32&-& 0.024 && 0.068&&\\
    &&&&&&&&&&&&&\\
    3:&&-&&-&&0.40&-&-&& 0.034&&\\
    &&&&&&&&&&&&&\\
    4:&&-&&-&&0.42&-&-&& 0.036&&\\
    &&&&&&&&&&&&&\\
    &&&&&&&&&&&&&\\
    $E^{{\rm 300 K}}_{SA}$ && $E_1$ && $E_2$ && $E_3$ & $\Delta_t$ & $E_{JT}$ \\
    &&&&&&&&&&&&&\\        
    ${\rm KCrF_3}$ && 1.42&& 1.65&& 1.82& 0.20&0.355&&&&&\\
    &&&&&&&&&&&&&\\    
    ${\rm NaCrF_3}$ && 1.40 && 1.66 &&1.78&0.12&0.350 &&&&&\\
    &&&&&&&&&&&&&\\
    &&&&&&&&&&&&&\\
    $E^{{\rm 2 K}}_{SA}$ && $E_1$ && $E_2$ && $E_3$ & $\Delta_t$ & $E_{JT}$ \\
    &&&&&&&&&&&&&\\        
    ${\rm KCrF_3}$ && 1.42&& 1.65&& 1.82& 0.20&0.355&&&&&\\
    &&&&&&&&&&&&&\\    
    ${\rm NaCrF_3}$ && 1.45 && 1.76 &&1.79&0.025&0.362 &&&&&\\
    &&&&&&&&&&&&&\\
    \hline
  \end{tabular}
\end{table}






\subsection{Temperature and magnetic field dependent OA Spectroscopy}
\label{sec:TDOA}

\fref{f8}  shows the  OA-spectra at  300 K  for $\mathrm{KCrF_3}$  and
$\mathrm{NaCrF_3}$, respectively.  The spectra are very similar in the
location of the observed bands. For $A{\rm CrF_3}$ there is no overlap
between  the  spin allowed  and  the  spin forbidden  transitions,  in
contrast to  the $A_x{\rm MnF}_{3+x}$ family.   The quintet-quintet or
spin-allowed ($E_{SA}$)  transitions (\fref{f0}  ($b$) and  ($c$)) lie
between 1.3 and 2 eV.  They are labeled here as $E_{SA}$ corresponding
to $E_1  ({}^5B_{1g}\to{}^5A_{1g})$, $E_2  ({}^5B_{1g}\to {}^5B_{2g})$
and $E_3  ({}^5B_{1g}\to{}^5E_g)$.  The  region between  2 and  2.6 eV
contains  the quintet-triplet,  or  spin-flip ($E_{SF}$)  transitions:
$E_4({}^5B_{1g}\to{}^3_aB_{1g})$                                   and
$E_5({}^5B_{1g}\to {}^3_bB_{1g})$.  The $E_{SF}$ bands for both phases
have similar  oscillating strengths despite ${\rm  NaCrF_3}$ having no
$\xi^{\pi}$ vertex angles.

We conducted temperature dependent OA measurements in the PPMS between
300 and 2 K and fitted the  $E_{SA}$ bands with a convolution of three
Gaussian functions, $\alpha t^2e^{(-\beta(E-E_i)^2)}$ (\fref{f2} ($a$)
and ($b$), respectively).  Inspection  of the temperature-dependent OA
measurements conducted in  the PPMS confirms the  patterns observed in
the preliminary  experiments shown in \fref{f8}.  The fitted positions
and oscillating intensities of the spin allowed bands at 300 K and 2 K
are given in \tref{T2} along with the $\Delta_t$ and $E_{JT}$ values.
The   $E_{SA}$  transitions   of  $\mathrm{KCrF_3}$   are  $E_1=1.42$,
$E_2=1.65$ and $E_3=1.82$ eV at 300 K.  At 2 K the positions of $E_1$,
$E_2$ and $E_3$ remain almost the same but their intensities decrease.
The  JT-stabilization  energy   $E_{JT}$  (Equation  \ref{eq:eA})  for
$\mathrm{KCrF_3}$ is  0.355 eV at  300 K and remains  nearly unchanged
down to 2 K.  The second  tetragonal parameter, $\Delta_t$, is 0.20 eV
at 300 K and also remains unchanged  down to 2 K.  The $E_{SA}$ energy
transitions  of  $\mathrm{NaCrF_3}$  are $E_1=1.40$,  $E_2=1.66$,  and
$E_3=1.785$ eV  at 300  K.  At  2 K  $E_1$ and  $E_2$ shift  to higher
energies  by 50  meV,  and  448 meV,  respectively.   This means  that
reduction of ion size at $A$-site  induces an increment in $E_{JT}$ of
12.5 meV  at 2  K which  in turn is  7.5 meV  higher than  $E_{JT}$ of
$\mathrm{KCrF_3}$.  The fit at 2 K shows that there is a total merging
of the $E_2$ and $E_3$ bands  as shown in \fref{f2} ($d$).  It follows
that the tetragonal parameter $\Delta_t$ of ${\rm NaCrF_3}$ disappears
and that $\Delta_e/\Delta_t\to \infty$ as $\Delta_t\to 0$.  This means
that  with  decreasing  temperature  contributions  to  $E_{JT}$  from
$\sigma$   bonding   become   stronger   in   ${\rm   NaCrF_3}$   than
$\mathrm{KCrF_3}$.  This effect is due  to the reduced symmetry caused
by  the ion  size reduction  at  the $A$-site.   The vibrational  mode
frequencies     $Q_{\vartheta}$      and     $Q_{\varepsilon}$     for
$\mathrm{ACrF_3}$ are calculated from the bond lengths by:

\begin{equation}
  \label{eq:e2}
  Q_{\vartheta}=(1/\sqrt{3}) (2l -m -s), \quad Q_{\varepsilon}=m -s
\end{equation}

The   values  are   given  in   \tref{T2}.   For   all  three   phases
$Q_{\vartheta}>>Q_{\varepsilon}$  which  suggests that  orbitals  with
$\ket{d_{3z^2-r^2}}$  symmetries  are   being  occupied.   The  normal
tetragonal mode $Q_{\vartheta}$ (i.e.   stretching of octahedral unit)
increases  on average  with decreasing  symmetry.  Since  $E_{JT}\sim$
$Q_{\vartheta}\sim  \rho$  (See  Equation  \ref{eq:e0})  we  see  that
$Q_{\vartheta} (\mathcal{I})>Q_{\vartheta} (\mathcal{M})>Q_{\vartheta}
(\mathcal{T})$ in  \tref{T2} thus the $E_{JT}$  for $\mathrm{NaCrF_3}$
is larger than  in $\mathrm{KCrF_3}$ indicating that  the reduction of
the $A$-ion size reinforces the JT-phenomenon.
\begin{figure}[t!]
  \flushleft
 \includegraphics[scale=.77]{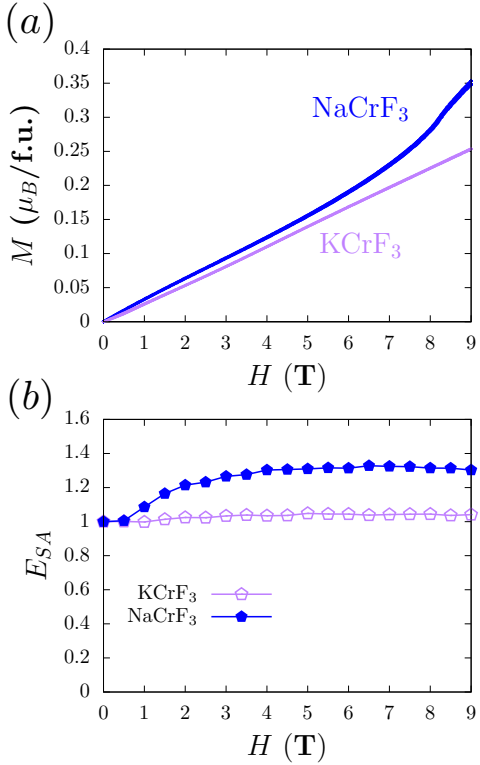}
 \caption{  ($a$) Isothermal  half-loop magnetization  curves magnetic
   field $(M (H))$ applied from 0 to 9 T and then back to 0 at 2 K for
   ${\rm KCrF_3}$ and ${\rm  NaCrF_3}$. ($b$) Magnetic field dependent
   OA  of  $E_{SA}$   transitions  (in  the  range   650-750  nm)  for
   $\mathrm{KCrF_3}$ and $\mathrm{NaCrF_3}$ at 2 K between 0-9 T.}
 \label{fig:f9}
\end{figure}

The intensities of  the $E_{SF}$ bands were taken  from the integrated
intensities of  the two peaks  for all temperatures.   \fref{f7} ($a$)
shows the fitted $E_{SF}$ intensities in  both phases as a function of
temperature,  superimposed on  the zero-field  cooling (ZFC)  magnetic
susceptibility, $\chi$, of both phases at an applied magnetic field of
1  T.  \fref{f7}  ($b$) shows  a close-up  below 60  K to  clarify the
behavior. The  temperature ($T_1$) in  $\mathrm{KCrF_3}$ corresponding
to the IC-AFM to C-AFM  transition (Ref.  \cite{PRB82}) and the Ne\'el
temperature  ($T_N$)  of $\mathrm{NaCrF_3}$  are  48.6  K and  21.5  K
respectively,      with     both      having     magnetic      moments
$\mu_{eff}\sim  4.47\mu_{B}$  following  the  spin-only  configuration
$S=2$  for $\mathrm{Cr^{2+}}$.  The two  samples obey  the Curie-Weiss
(CW) in different temperature  ranges: 300-100 K for $\mathrm{KCrF_3}$
and 300-24 K for  $\mathrm{NaCrF_3}$.  The Curie temperatures $\theta$
of  $\mathrm{KCrF_3}$   and  $\mathrm{NaCrF_3}$  are  1.7   and  -4  K
respectively (Refs. \cite{PRB82},\cite{ME}).

We observe a  direct magneto optic correlation  between the normalized
integrated intensities $E_{SF}$ of  $\mathrm{KCrF_3}$ and its magnetic
ordering as the  intensities of $E_{SF}$ start  decreasing smoothly at
$T_1$.  This is not observed for the case of ${\rm NaCrF_3}$ where the
$E_{SF}$ intensities decrease abruptly between 9 and 8 K.
The  disappearance of  the  $E_{SF}$  bands is  likely  to  lead to  a
thermochromic effect,  but we  have not  as yet  been able  to observe
this.
According to  our recent neutron  powder diffraction (NPD)  studies on
${\rm  NaCrF_3}$ (Ref.   \cite{ME})  ${\rm NaCrF_3}$  adopts a  canted
$A$-type  AFM ordering.   The refinements  were performed  using polar
coordinates   with   polar   and    azimuthal   degrees   of   freedom
($\vartheta,\varphi$).  The  polar angle $\varphi^\circ$  measures the
out-of-plane spin canting from  the $a_pc_p$-planes, and the azimuthal
angle $\vartheta^\circ$ the spin canting within the $a_pc_p$-planes.

\fref{f7} ($c$)  shows the comparison  of the $E_{SF}$  intensities of
${\rm  NaCrF_3}$ below  9 K  against the  temperature dependent  polar
angle $\varphi^\circ$  from the neutron powder  diffraction (NPD) data
(Ref \cite{ME}).   We observe that  below 9  K, as the  $E_{SF}$ bands
disappear, the  polar angles  of Cr$_1$ and  Cr$_3$, $\varphi_1^\circ$
and   $\varphi_3^\circ$   approach   the   same   value   (Note:   The
$\varphi^\circ$   components   of   the  magnetic   moments   of   the
${\rm  Cr^{2+}}$ ions  were  constrained in  the  NPD refinements  for
Cr$_2$ and Cr$_4$ as  follows: $\varphi_2^\circ=\varphi_1^\circ + 180$
and $\varphi_4^\circ=\varphi_3^\circ + 180$, so only $\varphi_1^\circ$
and $\varphi_3^\circ$ are quoted here).
  
After the  temperature reached  2 K, we  performed field  dependent OA
experiments for  both ${\rm  KCrF_3}$ and  ${\rm NaCrF_3}$  to explore
possible correlations with the field  dependent experiments $M (H)$ in
Ref.  \cite{ME}  which  metamagnetic  transitions  were  observed  for
${\rm  NaCrF_?}$.  Quick  inspection of  the field  dependent magnetic
measurements in \fref{f9} ($a$) shows that the plot for ${\rm KCrF_3}$
is linear while  that of ${\rm NaCrF_3}$ is not.  This agrees with our
previous   results  \cite{ME}.   This   indicates   the  presence   of
metamagnetism in  ${\rm NaCrF_3}$, but not  ${\rm KCrF_3}$.  \fref{f9}
($b$)  shows  the integrated  intensities  of  the $E_{SA}$  bands  of
$\mathrm{KCrF_3}$  and $\mathrm{NaCrF_3}$  as a  function of  magnetic
field in steps  of 0.5 T in the field  region between 0 - 9 T  at 2 K.
The  integrated  normalized  oscillating   strength  of  the  $E_{SA}$
transitions in $\mathrm{NaCrF_3}$ increases with magnetic field, while
for $\mathrm{KCrF_3}$, it remains constant.

\subsection{Discussion}
\label{sec:con}

The  JT-active  $A{\rm  CrF_3}$  systems,  where,  $A={\rm  K^+,Na^+}$
display  interesting  magneto-optic  and  structural  phenomena  under
external stimuli.   XRD reveals that  with decreasing ion size  at the
$A$-site,              $E_{JT}$              increases              as
$Q_{\vartheta}               (\mathcal{I})              >Q_{\vartheta}
(\mathcal{T})>Q_{\vartheta} (\mathcal{M})$.

Calculations of 
splitting parameters for  all phases of $A{\rm CrF_3}$  show that: (1)
the electron-ion JT  $E\otimes e$ coupling is essential for  the OO to
occur.  (2) A MIT was observed for $\mathcal{T}$ and $\mathcal{M}$ for
$U>1.5$ eV. This  indicates that for OO to  occur, JT-distortions must
already  be  in  place.   This  is  confirmed  with  the  substitution
${\rm K^+}\to{\rm Na^+}$.
(3)  At zero  electron-electron correlation  the tetragonal  splitting
parameter  $\Delta_e$  is zero  (ie.,  $\Delta_e\neq0$  for all  $U$).
Since the stabilization energy $E_{JT}=\Delta_e/4$, therefore a direct
correlation between  the results  of the  structural analysis  and the
calculations                       supports                       that
$E_{JT} (\mathcal{I})>E_{JT} (\mathcal{T,M})$.

As symmetry  is lowered  and the number  of $\xi^{\pi}$  decreases, so
does the  strength of the  magnetic interactions.  This  establishes a
magneto-structural correlation  between the $\xi^\circ$  vertex angles
and  the  magnetic  superexchange parameters.   The  calculated  super
exchange parameters  $J_{A}$ and $J_{B}$  show that by  decreasing the
lattice symmetry (and thereby the number of $\xi^{\pi}$ vertex angles)
the magnetic interactions reduce in strength regardless of $U$.
Thanks to  the fact  that the  $E_{SA}$ and $E_{SF}$  bands in  the OA
spectra of $A\mathrm{CrF_3}$ do not overlap, we can address the effect
of external stimuli on the two types of transition independently.


Temperature dependent OA-measurements  show that for $\mathrm{KCrF_3}$
the $E_{SA}$  energies are  not effected by  temperature in  the range
studied. For  $\mathrm{NaCrF_3}$ the $E_{SA}$ transitions  are shifted
to higher  energies with decreasing  temperature with $E_2$  and $E_3$
actually merging at 2 K.  This  means that $\Delta_t$ tends to becomes
zero,  which  indicates  that   electron-ion  couplings  arising  from
$\pi$-couplings   in   ${\rm   NaCrF_3}$   are   nearly   absent   and
$\Delta_e/\Delta_t \to \infty$.

The $E_1$ band in ${\rm NaCrF_3}$ shows a slight blue-shifting at 2 K,
which can  be interpreted as an  increase in $E_{JT}$ with  respect to
${\rm KCrF_3}$, again showing that reducing the size of the ion at the
$A$-site increases $E_{JT}$.  The  $\Delta_e$ values obtained from the
OA-measurements of $A\mathrm{CrF_3}$ lie  between $\Delta_e$ values of
$A'_3\mathrm{MnF_6}$ and $A'_2\mathrm{MnF_5}$ \cite{JCP}.  $\Delta_e$,
and  thus $E_{JT}$  for  $A\mathrm{CrF_3}$ do  not  follow the  linear
relation  between  $\Delta_e$  and  dimensionality  proposed  for  the
$A_x\mathrm{MnF_{3+x}}$ family.
  
The increase in $E_{SA}$ band intensity with increasing magnetic field
for $\mathrm{NaCrF_3}$ at 2 K is not observed for $\mathrm{KCrF_3}$ or
other related  ${\rm Cr^{2+}}$ fluorides,  e.g.  $\mathrm{Rb_2CrCl_4}$
in which the $E_{SF}$ transitions decline in intensity with increasing
magnetic field \cite{CrCl1,CrCl2}.
The  increase could  be  related  to the  orbital  structure and  spin
ordering  correlations  described  in the  Kugel-Khomskii  Hamiltonian
\cite{KK}.   However,  at this  point  we  cannot establish  a  direct
correlation  between  the  two  effects and  further  theoretical  and
experimental studies are necessary  for proper interpretation of these
results.

No significant changes in the $E_{SF}$ intensities are observed in the
OA-spectra  of ${\rm  KCrF_3}$  at the  $\mathcal{T}$-to-$\mathcal{M}$
transition.  Based  on the $A_x{\rm  MnF}_{3+x}$ family, in  which the
$E_{SF}$  transitions are  related to  $\xi^{\circ}$ and  thus to  the
magnetic  superexchange  interactions  $J_{A (B)}$,  we  expected  the
intensities  to decrease  as  the number  of  $\xi^{\pi}$ is  reduced.
Instead  of changing  at  the crystallographic  phase transition,  the
$E_{SF}$  intensities of  $\mathrm{KCrF_3}$  drop at  the IC-to  C-AFM
transition at $T_1=48.5$ K.
In  $\mathrm{NaCrF_3}$ the  $E_{SF}$  transition intensities  decrease
abruptly below 9 K  instead of at the Ne\'el temperature  of 21.5 K as
might  be  expected.  The   magnetic  susceptibility  measurements  of
$\mathrm{NaCrF_3}$ show a weak ferromagnetic upswing between 9 and 8 K
which coincides with the fall  in the $E_{SF}$ intensities.  The polar
components of the spins $\varphi^\circ$ in ${\rm NaCrF_3}$ established
by the NPD measurements describe  the component of the spins departing
from the canted AOO layers.  The $\varphi^\circ$ converge below 9 K in
a similar pattern  to the $E_{SF}$ intensities.  The  reduction of the
$E_{SF}$ at  $T_1$ in  ${\rm KCrF_3}$  and at 9  K in  ${\rm NaCrF_3}$
indicates that the intensity of $E_{SF}$ is controlled by the magnetic
structure  and not  the  crystal structure,  a direct  magneto-optical
correlation.   No  magneto-structural  correlations were  observed  at
standard  pressure  in  either   temperature  or  field  dependent  OA
experiments.

\section{Conclusions}
The  magneto-optical and  structural  properties of  fluoroperovskites
$\mathrm{ACrF_3}$  where $A  = {\rm  Na^+, K^+}$  display trends  that
contrast  strongly with  those reported  for the  $A_x{\rm MnF}_{3+x}$
family.
  
Analysis of  the $E_{SA}$ bands  in the  OA spectra (supported  by DFT
calculations and  XRD characterization) shows that  having the smaller
Na$^+$ ion  at the  $A$-site increases  the JT-  stabilization energy,
$E_{JT}$.
Field-dependent  OA measurements  of the  $E_{SA}$ in  ${\rm NaCrF_3}$
show magneto-optical  correlations absent in ${\rm  KCrF_3}$ and other
related  structures.   The collapse  of  the  $E_{SF}$ intensities  in
$A{\rm CrF_3}$ below  48.6 K is directly linked to  the local magnetic
interactions and not the crystal structure-  the more the order of the
structure  is reduced,  the  more significant  the local  interactions
based  on the  magnetic  properties of  individual ${\rm  CrF_6^{4-}}$
octahedra  become.   In  conclusion,  the ion  size  at  the  $A$-site
controls   the    physical   properties   of   the    $A{\rm   CrF_3}$
fluoroperovskites.  This, combined with  the reliable synthesis routes
we  have  developed  for  this  family  of  materials,  opens  up  the
possibility of discovering novel states  of matter by manipulating the
$A$-site.


\section{Aknowledgements}
\label{sec:akn}

The authors would  like to thank Prof.   Serena Margadonna (University
of  Swansea,  Swansea, UK)  for  granted  financial support  from  the
Norwegian  Research  Council  (Norges Forskningsr\aa  d  NFR)  through
project  214260.   We  acknowledge   use  of  the  Norwegian  national
infrastructure  for  X-ray  diffraction and  scattering  (NFR  project
number 208896).

\bibliographystyle{plain}
\bibliography{ACrF}

\end{document}